\documentclass[aps,prd,twocolumn,showpacs,preprintnumbers]{revtex4-1}
\usepackage{graphicx} 
\usepackage{dcolumn}  
\usepackage{amsmath}
\usepackage{epsfig}
\usepackage{multirow}
\usepackage{mciteplus}

\long\def\inst#1{\par\nobreak\kern 4pt\nobreak
    {\itshape #1}\par\vskip 10pt plus 3pt minus 3pt}

\graphicspath{{./figs/}}
\DeclareGraphicsExtensions{.pdf,.PDF,png,.PNG}

\usepackage{ifthen}  
\newboolean{uprightparticles}
\setboolean{uprightparticles}{false} 

\usepackage{xspace} 
\usepackage{upgreek}

\usepackage{relsize}
\def\babar{\mbox{\slshape B\kern-0.1em{\smaller A}\kern-0.1em
    B\kern-0.1em{\smaller A\kern-0.2em R}}\xspace}

\def\argus  {\mbox{ARGUS}\xspace}

\def\pep    {PEP-II}

\def\MagUp {\mbox{\em Mag\kern -0.05em Up}\xspace}

\ifthenelse{\boolean{uprightparticles}}%
{

 \def\Pmu         {\ensuremath{\upmu}\xspace}

 \def\Ppi         {\ensuremath{\uppi}\xspace}                 
                  
 \def\Prho        {\ensuremath{\uprho}\xspace}

 \def\PDelta      {\ensuremath{\Delta}\xspace}                 
 \def\PXi      {\ensuremath{\Xi}\xspace}                 
 \def\PLambda      {\ensuremath{\Lambda}\xspace}                 
 \def\PSigma      {\ensuremath{\Sigma}\xspace}                 
 \def\POmega      {\ensuremath{\Omega}\xspace}                 
 \def\PUpsilon      {\ensuremath{\Upsilon}\xspace}                 
 
 \def\PB      {\ensuremath{\mathrm{B}}\xspace}                 
                  
 \def\PD      {\ensuremath{\mathrm{D}}\xspace}

 \def\PK      {\ensuremath{\mathrm{K}}\xspace}

 \def\Pc      {\ensuremath{\mathrm{c}}\xspace}                 
                  
 \def\Pe      {\ensuremath{\mathrm{e}}\xspace}

 \def\Pi      {\ensuremath{\mathrm{i}}\xspace}

 \def\Pq      {\ensuremath{\mathrm{q}}\xspace}

}
{

 \def\Pmu         {\ensuremath{\mu}\xspace}

 \def\Ppi         {\ensuremath{\pi}\xspace}                 
                  
 \def\Prho        {\ensuremath{\rho}\xspace}

 \mathchardef\PDelta="7101
 \mathchardef\PXi="7104
 \mathchardef\PLambda="7103
 \mathchardef\PSigma="7106
 \mathchardef\POmega="710A
 \mathchardef\PUpsilon="7107
                  
 \def\PB      {\ensuremath{B}\xspace}                 
                  
 \def\PD      {\ensuremath{D}\xspace}

 \def\PK      {\ensuremath{K}\xspace}

 \def\Pc      {\ensuremath{c}\xspace}                 
                  
 \def\Pe      {\ensuremath{e}\xspace}

 \def\Pi      {\ensuremath{i}\xspace}

 \def\Pq      {\ensuremath{q}\xspace}

}

\makeatletter
\@ifundefined{ptsize}{
\newcommand{\miniscule}{\@setfontsize\miniscule{5}{6}}
}
{
\ifcase \@ptsize \relax
  \newcommand{\miniscule}{\@setfontsize\miniscule{4}{5}}
\or
  \newcommand{\miniscule}{\@setfontsize\miniscule{5}{6}}
\or
  \newcommand{\miniscule}{\@setfontsize\miniscule{5}{6}}
\fi
}
\makeatother

\DeclareRobustCommand{\optbar}[1]{\shortstack{{\miniscule (\rule[.5ex]{1.25em}{.18mm})}
  \\ [-.7ex] $#1$}}

\def\epem       {{\ensuremath{\Pe^+\Pe^-}}\xspace}

\def\mumu       {{\ensuremath{\Pmu^+\Pmu^-}}\xspace}
\def\mue        {\ensuremath{\mu^{\pm}e^{\mp}}\xspace}
\def\emu        {\ensuremath{e^{\pm}\mu^{\mp}}\xspace}

\def\ellm       {{\ensuremath{\ell^-}}\xspace}

\def\quark     {{\ensuremath{\Pq}}\xspace}
\def\quarkbar  {{\ensuremath{\overline \quark}}\xspace}
\def\qqbar     {{\ensuremath{\quark\quarkbar}}\xspace}

\def\cquark    {{\ensuremath{\Pc}}\xspace}
\def\cquarkbar {{\ensuremath{\overline \cquark}}\xspace}
\def\ccbar     {{\ensuremath{\cquark\cquarkbar}}\xspace}

\def\pion   {{\ensuremath{\Ppi}}\xspace}
\def\piz    {{\ensuremath{\pion^0}}\xspace}

\def\pip    {{\ensuremath{\pion^+}}\xspace}
\def\pim    {{\ensuremath{\pion^-}}\xspace}

\def\rhomeson {{\ensuremath{\Prho}}\xspace}
\def\rhoz     {{\ensuremath{\rhomeson^0}}\xspace}

\def\kaon    {{\ensuremath{\PK}}\xspace}
  \def\Kbar    {{\kern 0.2em\overline{\kern -0.2em \PK}{}}\xspace}

\def\KorKbar    {\kern 0.18em\optbar{\kern -0.18em K}{}\xspace}

\def\Kp      {{\ensuremath{\kaon^+}}\xspace}
\def\Km      {{\ensuremath{\kaon^-}}\xspace}

\def\KpKm     {\ensuremath{\Kp \kern -0.16em \Km}\xspace}
\def\KS      {{\ensuremath{\kaon^0_{\mathrm{ \scriptscriptstyle S}}}}\xspace}

\def\Kstarzb {{\ensuremath{\Kbar{}^{*0}}}\xspace}

  \def\Dbar    {{\kern 0.2em\overline{\kern -0.2em \PD}{}}\xspace}
\def\D       {{\ensuremath{\PD}}\xspace}

\def\DorDbar    {\kern 0.18em\optbar{\kern -0.18em D}{}\xspace}
\def\Dz      {{\ensuremath{\D^0}}\xspace}

\def\Dp      {{\ensuremath{\D^+}}\xspace}

\def\Dstarp  {{\ensuremath{\D^{*+}}}\xspace}

\def\B       {{\ensuremath{\PB}}\xspace}
\def\Bbar    {{\ensuremath{\kern 0.18em\overline{\kern -0.18em \PB}{}}}\xspace}

\def\BorBbar    {\kern 0.18em\optbar{\kern -0.18em B}{}\xspace}
\def\Bz      {{\ensuremath{\B^0}}\xspace}
\def\Bzb     {{\ensuremath{\Bbar{}^0}}\xspace}
\def\Bu      {{\ensuremath{\B^+}}\xspace}
\def\Bub     {{\ensuremath{\B^-}}\xspace}

\def\BpBm    {\ensuremath{\Bu {\kern -0.16em \Bub}}\xspace}
\def\BzBzb   {\ensuremath{\Bz {\kern -0.16em \Bzb}}\xspace}
\def\BB      {\ensuremath{\B\Bbar}\xspace}

  \def\Y#1S{\ensuremath{\PUpsilon{(#1S)}}\xspace}

\def\FourS {{\Y4S}}

\def\Lbar        {{\ensuremath{\kern 0.1em\overline{\kern -0.1em\PLambda}}}\xspace}
\def\LorLbar    {\kern 0.18em\optbar{\kern -0.18em \PLambda}{}\xspace}

\def\BF         {{\ensuremath{\mathcal{B}}}\xspace}

\def\BR         {\BF}
\def\calB         {\BF}

\def\to                 {\ensuremath{\rightarrow}\xspace}

\def\order   {{\ensuremath{\mathcal{O}}}\xspace}

\newcommand{\dm}{{\ensuremath{\Delta m}}\xspace}

\def\AT#1     {\ensuremath{A_{\mathrm{T}}^{#1}}\xspace}           

\def\C#1      {\ensuremath{\mathcal{C}_{#1}}\xspace}                       
\def\Cp#1     {\ensuremath{\mathcal{C}_{#1}^{'}}\xspace}                    
\def\Ceff#1   {\ensuremath{\mathcal{C}_{#1}^{\mathrm{(eff)}}}\xspace}        
\def\Cpeff#1  {\ensuremath{\mathcal{C}_{#1}^{'\mathrm{(eff)}}}\xspace}       
\def\Ope#1    {\ensuremath{\mathcal{O}_{#1}}\xspace}                       
\def\Opep#1   {\ensuremath{\mathcal{O}_{#1}^{'}}\xspace}                    

\newcommand{\tev}{\ifthenelse{\boolean{inbibliography}}{\ensuremath{~T\kern -0.05em eV}\xspace}{\ensuremath{\mathrm{\,Te\kern -0.1em V}}}\xspace}
\newcommand{\gev}{\ensuremath{\mathrm{\,Ge\kern -0.1em V}}\xspace}
\newcommand{\mev}{\ensuremath{\mathrm{\,Me\kern -0.1em V}}\xspace}
\newcommand{\kev}{\ensuremath{\mathrm{\,ke\kern -0.1em V}}\xspace}
\newcommand{\ev}{\ensuremath{\mathrm{\,e\kern -0.1em V}}\xspace}
\newcommand{\gevc}{\ensuremath{{\mathrm{\,Ge\kern -0.1em V\!/}c}}\xspace}
\newcommand{\mevc}{\ensuremath{{\mathrm{\,Me\kern -0.1em V\!/}c}}\xspace}
\newcommand{\gevcc}{\ensuremath{{\mathrm{\,Ge\kern -0.1em V\!/}c^2}}\xspace}
\newcommand{\gevgevcccc}{\ensuremath{{\mathrm{\,Ge\kern -0.1em V^2\!/}c^4}}\xspace}
\newcommand{\mevcc}{\ensuremath{{\mathrm{\,Me\kern -0.1em V\!/}c^2}}\xspace}

\def\cm   {\ensuremath{\mathrm{ \,cm}}\xspace}

\def\invfb   {\ensuremath{\mbox{\,fb}^{-1}}\xspace}

\def\order{{\ensuremath{\mathcal{O}}}\xspace}
\newcommand{\chisq}{\ensuremath{\chi^2}\xspace}

\def\gsim{{~\raise.15em\hbox{$>$}\kern-.85em
          \lower.35em\hbox{$\sim$}~}\xspace}
\def\lsim{{~\raise.15em\hbox{$<$}\kern-.85em
          \lower.35em\hbox{$\sim$}~}\xspace}

\newcommand{\lum} {\ensuremath{\mathcal{L}}\xspace}

\def\EVTGEN     {\mbox{\tt EvtGen}\xspace} 

\def\geantfour  {\mbox{\tt GEANT 4}\xspace} 

\def\PHOTOS     {\mbox{\tt PHOTOS}\xspace} 

\def\jetset {\mbox{\tt Jetset\xspace}}
\def\tauola {\mbox{\tt TAUOLA\xspace}}
\def\kktwof {\mbox{\tt KK2F\xspace}}
\def\afkqed {\mbox{\tt AfkQed\xspace}}

\def\tell1  {TELL1\xspace}
\def\ukl1   {UKL1\xspace}

\newcommand{\eg}{\mbox{\itshape e.g.}\xspace}
\newcommand{\ie}{\mbox{\itshape i.e.}\xspace}

\def\onreslumi  {\ensuremath{424.3\pm1.8}}

\def\usedlumi   {\ensuremath{39.3\pm0.2}}  
\def\totallumiNoerr  {\ensuremath{468.2}}
\def\totallumi  {\ensuremath{\totallumiNoerr\pm2.0}}

\def\nsig       {\ensuremath{N_{\rm sig}}}
\def\nnorm      {\ensuremath{N_{\rm norm}}}

\def\DzToKPiee   {\ensuremath{\Dz\to\Km\pip\epem}}

\def\DpToXpll   {\ensuremath{\Dp\to X^{+}\ell^{\prime+}\ellm}}

\def\KKPiPi      {\ensuremath{\Km\Kp\pip\pim}}
\def\DzToKKPiPi  {\ensuremath{\Dz\to\KKPiPi}}

\def\KPiPiPi      {\ensuremath{\Km\pip\pip\pim}}
\def\DzToKPiPiPi  {\ensuremath{\Dz\to\KPiPiPi}}

\def\PiPiPiPi      {\ensuremath{\pim\pip\pip\pim}}
\def\DzToPiPiPiPi  {\ensuremath{\Dz\to\PiPiPiPi}}

\def\DzToKPi     {\ensuremath{\Dz\to\Km\pip}}

\def\multiKPiPiPi     {\ensuremath{3.6}}
\def\multiKKPiPi      {\ensuremath{2.4}}
\def\multiPiPiPiPi    {\ensuremath{4.4}}

\def\BFDzToKKPiPi   {\ensuremath{3.51\pm0.18\pm0.18}} 
\def\BFDzToKPiPiPi  {\ensuremath{3.98\pm0.08\pm0.10}} 
\def\BFDzToPiPiPiPi {\ensuremath{4.12\pm0.13\pm0.16}} 

\def\BFDzToKPi      {\ensuremath{3.95\pm0.03}} 

\def\effKPiNoErr       {\ensuremath{27.4}}
\def\effKKPiPiNoErr    {\ensuremath{19.2}}
\def\effKPiPiPiNoErr   {\ensuremath{20.1}}
\def\effPiPiPiPiNoErr  {\ensuremath{24.7}}

\def\effKPi       {\ensuremath{\effKPiNoErr\pm0.2}}
\def\effKKPiPi    {\ensuremath{\effKKPiPiNoErr\pm0.2}}
\def\effKPiPiPi   {\ensuremath{\effKPiPiPiNoErr\pm0.2}}
\def\effPiPiPiPi  {\ensuremath{\effPiPiPiPiNoErr\pm0.2}}

\def\yieldKPi       {\ensuremath{1\,881\,950\pm1380}} 
\def\yieldKKPiPi    {\ensuremath{8480\pm110}}       
\def\yieldKPiPiPi   {\ensuremath{260\,870\pm520}}     
\def\yieldPiPiPiPi  {\ensuremath{28\,470\pm220}}      

\def\systPiPiPiPi    {\ensuremath{6.8}}
\def\systKPiPiPi     {\ensuremath{4.7}}
\def\systKKPiPi      {\ensuremath{6.6}}

\def\DzTohheeOS    {\ensuremath{\Dz\to h^{\prime -}h^{+}\epem}}
\def\DzTohhmumuOS  {\ensuremath{\Dz\to h^{\prime -}h^{+}\mumu}}

\def\DzToXzemu  {\ensuremath{\Dz\to X^{0}\emu}}

\def\DzTohhemuOS   {\ensuremath{\Dz\to h^{\prime -} h^{+}e^{\pm}\mu^{\mp}}}

\def\DzToRhoemuOS    {\ensuremath{\Dz\to\rhoz\emu}}
\def\DzToKSemuOS     {\ensuremath{\Dz\to\KS\emu}}

\def\DzToPhiemuOS    {\ensuremath{\Dz\to\phi\,\emu}}
\def\DzToKstaremuOS  {\ensuremath{\Dz\to\Kstarzb\emu}}
\def\DzToPizemuOS    {\ensuremath{\Dz\to\piz\emu}}
\def\DzToOmegaemuOS  {\ensuremath{\Dz\to\omega\emu}}
\def\DzToEtaggemuOS  {\ensuremath{\Dz\to\eta\emu}}

\def\DzToRhoemuOSsub    {\ensuremath{\Dz\to\rhoz(\to\pip\pim)\emu}}
\def\DzToKSemuOSsub     {\ensuremath{\Dz\to\KS(\to\pip\pim)\emu}}
\def\DzToPhiemuOSsub    {\ensuremath{\Dz\to\phi(\to\Kp\Km)\emu}}
\def\DzToKstaremuOSsub  {\ensuremath{\Dz\to\Kstarzb(\to\Km\pip)\emu}}

\def\DzToEtaggemuOSsub  {\ensuremath{\Dz\to\eta(\to\gamma\gamma)\emu}}

\def\DzToEtaThreepiemuOSsub {\ensuremath{\Dz\to\eta(\to\pip\pim\piz)\emu}}

\def\DstarXp      {\ensuremath{x_p=p^{*}(\Dstarp)/\sqrt{E_{\epem}^{*2}-m^2(\Dstarp)}}}

\def\effPizemuOSNoErr        {\ensuremath{2.15}}
\def\effKSemuOSNoErr         {\ensuremath{3.01}}
\def\effKstaremuOSNoErr      {\ensuremath{2.31}}
\def\effRhoemuOSNoErr        {\ensuremath{2.10}}
\def\effPhiemuOSNoErr        {\ensuremath{3.43}}
\def\effOmegaemuOSNoErr      {\ensuremath{1.46}}
\def\effEtaggemuOSNoErr      {\ensuremath{2.96}}
\def\effEtaThreepiemuOSNoErr {\ensuremath{2.46}}

\def\effPizemuOS        {\ensuremath{\effPizemuOSNoErr\pm0.03}}
\def\effKSemuOS         {\ensuremath{\effKSemuOSNoErr\pm0.04}}
\def\effKstaremuOS      {\ensuremath{\effKstaremuOSNoErr\pm0.03}}
\def\effRhoemuOS        {\ensuremath{\effRhoemuOSNoErr\pm0.03}}
\def\effPhiemuOS        {\ensuremath{\effPhiemuOSNoErr\pm0.04}}
\def\effOmegaemuOS      {\ensuremath{\effOmegaemuOSNoErr\pm0.03}}
\def\effEtaggemuOS      {\ensuremath{\effEtaggemuOSNoErr\pm0.04}}
\def\effEtaThreepiemuOS {\ensuremath{\effEtaThreepiemuOSNoErr\pm0.04}}

\def\obsPizemuOS        {\ensuremath{-0.3 \pm 2.0\pm 0.9}}
\def\obsKSemuOS         {\ensuremath{0.7  \pm 1.7\pm 0.7}}
\def\obsKstaremuOS      {\ensuremath{0.8  \pm 1.8\pm 0.8}}
\def\obsRhoemuOS        {\ensuremath{-0.7 \pm 1.7\pm 0.4}}
\def\obsPhiemuOS        {\ensuremath{0.0 \pm 1.4\pm 0.3}}
\def\obsOmegaemuOS      {\ensuremath{0.4 \pm 2.3\pm 0.5}}
\def\obsEtaggemuOS      {\ensuremath{1.6 \pm 2.3\pm 0.5}}
\def\obsEtaThreepiemuOS {\ensuremath{0.0 \pm 2.8\pm 0.7}}

\def\ulactPizemuOS       {\ensuremath{8.0}}
\def\ulactKSemuOS        {\ensuremath{8.6}}
\def\ulactKstaremuOS     {\ensuremath{12.4}}
\def\ulactRhoemuOS       {\ensuremath{5.0}}
\def\ulactPhiemuOS       {\ensuremath{5.1}}
\def\ulactOmegaemuOS     {\ensuremath{17.1}}
\def\ulactEtaggemuOS     {\ensuremath{24.0}}
\def\ulactEtaThreeemuOS  {\ensuremath{42.8}}
\def\ulactEtaemuOS       {\ensuremath{22.5}}

\def\systBFPizemuOS        {\ensuremath{2.3}}
\def\systBFKSemuOS         {\ensuremath{1.9}}
\def\systBFKstaremuOS      {\ensuremath{2.6}}
\def\systBFRhoemuOS        {\ensuremath{1.0}}
\def\systBFPhiemuOS        {\ensuremath{0.9}}
\def\systBFOmegaemuOS      {\ensuremath{1.9}}
\def\systBFEtaggemuOS      {\ensuremath{2.4}}
\def\systBFEtaThreepiemuOS {\ensuremath{6.0}}
\def\systBFEtaemuOS        {\ensuremath{2.3}}

\def\bfactPizemuOS        {\ensuremath{-0.6 \pm 4.8  \pm\systBFPizemuOS}}
\def\bfactKSemuOS         {\ensuremath{1.9 \pm 4.6  \pm\systBFKSemuOS}}
\def\bfactKstaremuOS      {\ensuremath{2.8 \pm 6.1  \pm\systBFKstaremuOS}}
\def\bfactRhoemuOS        {\ensuremath{-1.8 \pm 4.4  \pm\systBFRhoemuOS}}
\def\bfactPhiemuOS        {\ensuremath{0.1 \pm 3.8  \pm\systBFPhiemuOS}}
\def\bfactOmegaemuOS      {\ensuremath{1.8 \pm 9.5  \pm\systBFOmegaemuOS}}
\def\bfactEtaggemuOS      {\ensuremath{7.0 \pm 10.5 \pm\systBFEtaggemuOS}}
\def\bfactEtaThreepiemuOS {\ensuremath{0.4 \pm 25.8 \pm\systBFEtaThreepiemuOS}}
\def\bfactEtaemuOS        {\ensuremath{6.1 \pm 9.7   \pm\systBFEtaemuOS}}

\newcommand{\BABARPubYear}    {20}
\newcommand{\BABARPubNumber}  {001}
\newcommand{\SLACPubNumber} {17524}

\begin{document}
\preprint{\babar-PUB-\BABARPubYear/\BABARPubNumber} 
\preprint{SLAC-PUB-\SLACPubNumber}

\title{{\Large \bf \boldmath Search for Lepton-Flavor-Violating Decays $D^{0}\rightarrow X^{0}e^{\pm}\mu^{\mp}$}}

\author{J.~P.~Lees}
\author{V.~Poireau}
\author{V.~Tisserand}
\affiliation{Laboratoire d'Annecy-le-Vieux de Physique des Particules (LAPP), Universit\'e de Savoie, CNRS/IN2P3,  F-74941 Annecy-Le-Vieux, France}
\author{E.~Grauges}
\affiliation{Universitat de Barcelona, Facultat de Fisica, Departament ECM, E-08028 Barcelona, Spain }
\author{A.~Palano}
\affiliation{INFN Sezione di Bari and Dipartimento di Fisica, Universit\`a di Bari, I-70126 Bari, Italy }
\author{G.~Eigen}
\affiliation{University of Bergen, Institute of Physics, N-5007 Bergen, Norway }
\author{D.~N.~Brown}
\author{Yu.~G.~Kolomensky}
\affiliation{Lawrence Berkeley National Laboratory and University of California, Berkeley, California 94720, USA }
\author{M.~Fritsch}
\author{H.~Koch}
\author{T.~Schroeder}
\affiliation{Ruhr Universit\"at Bochum, Institut f\"ur Experimentalphysik 1, D-44780 Bochum, Germany }
% ...
\collaboration{The \babar\ Collaboration}
\noaffiliation

\begin{abstract}
\noindent We present a search for seven lepton-flavor-violating
neutral charm meson decays of the type $D^{0}\rightarrow X^{0} e^{\pm}
\mu^{\mp}$, where $X^{0}$ represents a $\pi^{0}$, $K^{0}_{\rm S}$,
${{\kern 0.2em\overline{\kern -0.2em K}{}}\xspace}{}^{*0}$,
$\rho^{0}$, $\phi$, $\omega$, or $\eta$ meson. The analysis is based
on $468$~{\ensuremath{\mbox{\,fb}^{-1}}} of $e^+e^-$ annihilation data
collected at or close to the {\ensuremath{{\it \Upsilon}(4S)}\xspace}
resonance with the \babar\ detector at the SLAC National Accelerator
Laboratory. No significant signals are observed, and we establish 90\%
confidence level upper limits on the branching fractions in the range
$(5.0 - 22.5)\times 10^{-7}$. The limits are between 1 and 2 orders of
magnitude more stringent than previous measurements.
\end{abstract}

\pacs{13.25.Ft, 11.30.Fs}
  
\maketitle

\section{Introduction}

Lepton-flavor-conserving charm decays such as $D\to X\epem$ or $D\to
X\mumu$, where $X$ is a meson, can occur in the standard model (SM)
through short-distance~\cite{Paul:2011ar,Schwartz:1993aa} and
long-distance~\cite{Schwartz:1993aa} processes, which have branching
fractions of order $\order(10^{-9})$ and $\order(10^{-6})$,
respectively.  In contrast, the lepton-flavor-violating (LFV) neutral
charm decays \DzToXzemu, where $X^0$ is a neutral meson, are
effectively forbidden in the SM because they can occur only through
lepton-flavor mixing~\cite{GUADAGNOLI201554} and are therefore
suppressed to the order $\order(10^{-50})$. As such, the
decays \DzToXzemu\ should not be visible with current data
samples. However, new-physics models, such as those involving Majorana
neutrinos, leptoquarks, and two-Higgs doublets, allow for lepton number
and lepton flavor to be
violated~\cite{deBoer:2015boa,Fajfer:2015mia,Atre:2009aa,Yuan:2013yba,1674-1137-39-1-013101}. Some
models make predictions for, or use constraints from, three-body
decays of the form $\D\to Xl'l$ or $\B\to
Xl'l$, where $\ell$ and $\ell^{\prime}$ represent an
electron or muon~\cite{Paul:2011ar,Paul:2012ab,Burdman:2001tf,Fajfer:2005ke,PhysRevD.76.074010,Atre:2009aa,Yuan:2013yba}. Most
recent theoretical work has targeted the charged charm decays
\DpToXpll. For example, Ref.~\cite{deBoer:2015boa} estimates that
$\BR(\Dp\to\pip\mue)$ can be as large as $2\times 10^{-6}$ for certain
leptoquark couplings. Some models that consider LFV and
lepton-number-violating (LNV) four-body
charm decays, with two leptons and two hadrons in the final state,
predict branching fractions up to $\order(10^{-5})$, approaching those
accessible with current
data~\cite{Atre:2009aa,Yuan:2013yba,1674-1137-39-1-013101}.

The branching fractions $\BR(\DzTohhmumuOS)$, where $h^{\prime}$ and
$h$ represent a $K$ or $\pi$ meson, and $\BR(\DzToKPiee)$ have
recently been measured to be $\order(10^{-7})$ to
$\order(10^{-6})$~\cite{LHCb-PAPER-2015-043,LHCb-PAPER-2017-019,Lees:2018zz},
compatible with SM
predictions~\cite{Cappiello:2012vg,Gudrun:035041}. The branching
fractions for the decays $\Dz\to X^0\epem$ and $\Dz\to X^0\mumu$ have
not yet been measured. However 90\% confidence level (C.L.) upper
limits on the branching fractions do exist and are in the range
$(0.3-10)\times10^{-5}$ for $\Dz\to X^0\epem$ and
$(3.2-53)\times10^{-5}$ for $\Dz\to
X^0\mumu$~\cite{Kodama:1995ia,Freyberger:1996it,Aitala:2000kk,PhysRevD.97.072015}. It
is likely that one or more of these decays are a major contributor to
the branching fractions of the decays \DzTohheeOS\ or \DzTohhmumuOS,
as long-distance processes are predicted to be
dominant~\cite{Schwartz:1993aa}, and published distributions of the
invariant masses $m(h^{\prime -}h^{+})$ for \DzTohhmumuOS\ and
\DzToKPiee\ indicate large yields near some of the $X^0$ invariant
masses~\cite{LHCb-PAPER-2015-043,LHCb-PAPER-2017-019,Lees:2018zz}.

The most stringent existing upper limits on the branching fractions
for the LFV four-body decays of the type \DzTohhemuOS\ are in the
range $(11.0-19.0)\times 10^{-7}$ at the 90\% confidence
level~\cite{Lees:2019zz}. For the LFV decays \DzToXzemu, where $X^{0}$
is an intermediate resonance meson decaying to $h^{\prime -} h^{+}$,
$\pip\pim\piz$ or $\gamma\gamma$, the 90\% C.L. limits are in the
range $(3.4-118)\times
10^{-5}$~\cite{Freyberger:1996it,Aitala:2000kk,PDG2019}.  For the
\DzToXzemu\ decays with the same final state as the
\DzTohhemuOS\ decays  (\DzToKSemuOSsub, \DzToRhoemuOSsub,
\DzToKstaremuOSsub, and \DzToPhiemuOSsub), the current
\DzToXzemu\ branching fraction upper limits, which are in the range
$(3.4-8.3)\times
10^{-5}$~\cite{Freyberger:1996it,Aitala:2000kk,PDG2019}, are
approximately 20 times less stringent than the \DzTohhemuOS\ limits
reported in Ref.~\cite{Lees:2019zz}.

In this report we present a search for seven \DzToXzemu\ LFV decays,
where $X^{0}$ represents a \piz, \KS, \Kstarzb, $\rho^{0}$, $\phi$,
$\omega$, or $\eta$ meson, with data recorded with the
\babar\ detector at the \pep\ asymmetric-energy $\epem$ collider
operated at the SLAC National Accelerator Laboratory. The intermediate
mesons $X^{0}$ are reconstructed through the decays
$\piz\to\gamma\gamma$, $\KS\to\pip\pim$, $\Kstarzb\to\Km\pip$,
$\rhoz\to\pip\pim$, $\phi\to\KpKm$, $\omega\to\pip\pim\piz$,
$\eta\to\pip\pim\piz$, and $\eta\to\gamma\gamma$.  The branching
fractions for the signal modes are measured relative to the
normalization decays \DzToPiPiPiPi\ (for $X^{0} = \KS, \rhoz,
\omega$), \DzToKPiPiPi\ ($X^0=\Kstarzb$), and
\DzToKKPiPi\ ($X^0=\phi$). For $X^{0} = \piz$ or $\eta$, the
normalization mode \DzToKPiPiPi\ is used as it has the smallest
branching fraction uncertainty~\cite{PDG2019} and the largest number
of reconstructed candidates of the three normalization modes.
Although decays of the type $\Dz\to X^0 h^{\prime -} h^{+}$ have
momentum distributions that more closely follow those of the signal decays
under study, they suffer from smaller branching fractions, greater
uncertainties on their branching fractions, and reduced reconstruction
efficiencies relative to the three chosen normalization modes.

The \Dz\ mesons are identified using the decay $\Dstarp\to\Dz\pip$
produced in $\epem\to\ccbar$ events. Although \Dz\ mesons are also
produced via other processes, the use of this decay chain
increases the purity of the \Dz\ samples at the cost of a smaller
number of reconstructed \Dz\ mesons.

\section{The \babar\ detector and data set}
\label{data}

The \babar\ detector is described in detail in
Refs.~\cite{Aubert:2001tu,TheBABAR:2013jta}. Charged particles are
reconstructed as tracks with a five-layer silicon vertex detector and
a 40-layer drift chamber inside a $1.5\,$T solenoidal magnet. An
electromagnetic calorimeter comprised of 6580 CsI(Tl) crystals is used
to identify and measure the energies of electrons, positrons, muons,
and photons. A ring-imaging Cherenkov detector is used to identify
charged hadrons and to provide additional lepton identification
information. Muons are primarily identified with an instrumented
magnetic-flux return.

The data sample corresponds to 424\invfb\ of \epem\ collisions
collected at the center-of-mass (c.m.) energy of the \FourS\ resonance
(10.58\gev, on peak) and an additional 44\invfb\ of data collected
0.04~\gev\ below the \FourS\ resonance (off peak)~\cite{Lees:2013rw}.

Monte Carlo (MC) simulation is used to investigate sources of
background contamination and evaluate selection
efficiencies. Simulated events are also used to validate the selection
procedure and for studies of systematic effects.  The signal and
normalization channels are simulated with the
\EVTGEN\ package~\cite{Lange:2001uf}. We generate the signal channel
decays uniformly throughout the three-body phase space, while the
normalization modes include two-body and three-body intermediate
resonances, as well as nonresonant decays. We also generate
$\epem\to\qqbar$ ($q=u,d,s,c$), Bhabha and \mumu\ pairs (collectively
referred to as QED events), and \BB\ background, using a combination
of the \EVTGEN, \jetset~\cite{Sjostrand:1993yb},
\kktwof~\cite{Ward:2002qq}, \afkqed~\cite{Czyz:2000wh}, and
\tauola~\cite{Davidson:2010rw} generators, where appropriate. The
background samples are produced with an integrated luminosity
approximately 6 times that of the data. Final-state radiation is
generated using \PHOTOS~\cite{Golonka:2005pn}. The detector response
is simulated with
\geantfour~\cite{Agostinelli:2002hh,Allison:2006ve}. All simulated
events are reconstructed in the same manner as the data.

\section{Event Selection}
\label{selection}

In the following, unless otherwise noted, all observables are
evaluated in the laboratory frame.  In order to optimize the event
reconstruction, candidate selection criteria, multivariate analysis
training, and fit procedure, a rectangular area in the $m(\Dz)$ versus
$\dm = m(\Dstarp)-m(\Dz)$ plane is defined, where $m(\Dstarp)$ and
$m(\Dz)$ are the reconstructed masses of the \Dstarp\ and
\Dz\ candidates, respectively. This region is kept hidden (blinded) in
data until the analysis steps are finalized. The hidden region is
approximately 3 times the root mean square (RMS) width of the \dm\ and
$m(\Dz)$ resolutions. Its \dm\ region is $0.1447<\dm<0.1462\gevcc$ for
all modes. The $m(\Dz)$ signal peak distribution is asymmetric due to
bremsstrahlung emission, with the left-side RMS width typically $1$ to
$2$\mevcc wider than the right-side. The $m(\Dz)$ RMS widths vary
between $5$ and $21$\mevcc, depending on the signal mode.

Particle identification (PID) criteria are applied to all charged
daughter tracks of the intermediate meson $X^{0}$ decays.  The charged
pions and kaons are identified by measurements of their energy loss in
the tracking detectors, and the number of photons and the Cherenkov
angle recorded in the ring-imaging Cherenkov detector. These
measurements are combined with information from the 
electromagnetic calorimeter and the muon detector to identify
electrons and muons~\cite{Aubert:2001tu,TheBABAR:2013jta}. Photons
are detected and their energies are measured in the electromagnetic
calorimeter. For \DzToPhiemuOS, the PID requirement on the kaons
from the $\phi$ meson decay is relaxed compared to the single-kaon
modes.  This increases the reconstruction efficiency for this signal
mode, with little increase in backgrounds or misidentified
candidates. The muon PID requirement depends on the signal mode, with 
tighter requirements imposed for modes with more charged pions in the
final state. The PID efficiency depends on the track momentum, and is
in the range $0.87-0.92$ for electrons, $0.60-0.95$ for muons,
$0.86-0.98$ for pions, and $0.84-0.92$ for kaons. The
misidentification probability~\cite{Bevan:2014iga}, defined as the
probability that particles are identified as one flavor (\eg\ muon)
that are in reality of a different flavor (\ie\ not a muon), is
typically less than $0.03$ for all selection criteria, except for the
pion selection criteria, where the muon misidentification rate can be
as high as $0.35$ at low momentum.

We select events that have at least five charged tracks, except for
\DzToPizemuOS\ and \DzToEtaggemuOSsub, which must have at least three.
Two or more of the tracks must be identified as leptons. The
separation along the beam axis between the two leptons at their
distance of closest approach to the beam line is required to be less than
0.2\cm. The leptons must have opposite charges, and their momenta must
be greater than 0.3\gevc. Electrons and
positrons from photon conversions are rejected by removing
electron-positron pairs with an invariant mass less than 0.03\gevcc
and a production vertex more than 2\cm\ from the beam axis.

The minimum photon energy in a signal decay is required to be greater
than 0.025\gev. For the decays \DzToPizemuOS\ and \DzToEtaggemuOSsub,
the momentum of the \piz\ or $\eta$ must be greater than 0.4\gevc and
the energy of each photon from the \piz\ must be greater than 0.045\gev. The
reconstructed \piz\ invariant mass for all signal decays is required
to be between 120 and 160\mevcc. 

The reconstructed invariant masses of the \piz, \KS, \Kstarzb,
$\rhoz$, $\phi$, and $\omega$ candidates are required to be within 19,
9, 76, 240, 20, and 34\mevcc, of their nominal mass~\cite{PDG2019},
respectively. For the decays $\eta\to\gamma\gamma$ and
$\eta\to\pip\pim\piz$, the invariant mass of the $\eta$ candidates
must be within 47 and 35\mevcc\ of the $\eta$ nominal mass,
respectively. These ranges are equivalent to 3 times the reconstructed
RMS widths.

Candidate \Dz\ mesons for the signal modes are formed from the
electron or positron, muon or antimuon, and intermediate resonance candidates. For
the normalization modes, the \Dz\ candidate is formed from four
charged tracks.  Particle identification is applied to all charged
tracks and the \Dz\ candidates are reconstructed with the appropriate
charged-track mass hypotheses for both the signal and normalization
decays. The tracks are required to form a good-quality vertex with a
\chisq\ probability for the vertex fit greater than 0.005. For the decay \DzToKSemuOS, the
\KS\ must have a transverse flight distance from the \Dz\ decay vertex
greater than 0.2\cm. A bremsstrahlung energy recovery algorithm is applied to electrons and
positrons, in which the energy of photon showers that are within a
small angle (35 mrad in polar angle and 50 mrad in
azimuth~\cite{Aubert:2001tu}) with respect to the tangent of the
initial electron or positron direction is added to the energy of the
electron or positron candidate.  For the normalization modes, the
reconstructed \Dz\ meson mass is required to be in the range
$1.81<m(\Dz)<1.91\gevcc$, while for the signal modes, $m(\Dz)$ must be
in the hidden $m(\Dz)$ range defined above. 

The candidate \Dstarp\ is formed by combining the \Dz\ candidate with
a charged pion having a momentum greater than
$0.1\gevc$. For the normalization mode \DzToKPiPiPi, this pion is
required to have a charge opposite that of the kaon. The pion and
\Dz\ candidate are subject to a vertex fit, with the \Dz\ mass
constrained to its known value~\cite{PDG2019} and the requirement that
the \Dz\ meson and the pion originate from the
beam spot~\cite{Hulsbergen:2005pu}. The \chisq\ probability of the fit
is required to be greater than 0.005. After the application of the
\Dstarp\ vertex fit, the \Dz\ candidate momentum in the c.m.\ system
$p^{\ast}(\Dz)$ must be greater than 2.4\gevc. For the normalization
modes, the mass difference \dm\ is required to be
$0.143<\dm<0.148\gevcc$, while for the signal modes the range is
$0.1395<\dm<0.1610\gevcc$. The extended \dm\ range for the signal
modes provides greater stability when fitting the background
distributions.

 The requirement on the number of charged tracks strongly suppresses
 backgrounds from QED processes. The $p^{\ast}(\Dz)$ criterion removes most
 sources of combinatorial background, as well as charm hadrons produced
 in \B\ decays, which are kinematically limited to $p^{\ast}(\Dz) \lesssim
 2.2\gevc$~\cite{PhysRevD.69.111104}. 

Simulated samples indicate that the remaining background arises from
$\epem\to\ccbar$ events in which charged tracks and neutral particles
can either be lost or selected from elsewhere in the event to form a
\Dz\ candidate. To reject this background, a multivariate selection
based on a Boosted Decision Tree (BDT) discriminant is applied to the
signal modes~\cite{AdaBoost}.  A common set of
eight input observables is used for all modes: the momenta of the
electron or positron, muon or antimuon, and reconstructed intermediate
meson; the momentum of the lowest-momentum charged track or photon
from the $X^0$ candidate; the maximum angle between the direction of
\Dz\ daughters and the \Dz\ direction; the total energy of all charged
tracks and photons in the event, normalized to the beam energy; the
ratio \DstarXp, where $p^*(\Dstarp)$ is the c.m. momentum of the
\Dstarp\ candidate and $E^*_{\epem}$ is the c.m. beam energy; and the
reconstructed mass of the intermediate meson. Three additional input
observables are used for the \Dz\ decays with $\omega$ or $\eta$
decaying to $\pip\pim\piz$: the momentum and reconstructed mass of the
\piz\ candidate, and the energy of the lowest-energy photon from the
\piz. The discriminant is trained and tested independently for each
signal mode, using simulated samples for the signal modes, and
ensembles of data outside the hidden region and $\epem\to\ccbar$
simulated samples for the background. Depending on the signal mode,
the requirement on the discriminant output accepts between 70\% to
90\% of the simulated signal sample while rejecting between 50\% to
90\% of the background.

The cross feed to one signal mode from any other signal modes is
estimated from simulated samples to be less than $4\%$ in all cases,
and typically less than $1\%$, assuming equal branching fractions for
all signal modes. The cross feed to a specific normalization mode from
the other two normalization modes is predicted from simulation to be
less than $0.7\%$, where the branching fractions are taken from
Ref.~\cite{PDG2019}.  In the data, no events with reconstructed
normalization decays contain reconstructed signal decays.

From the data, we find that the fraction of normalization mode events
with more than one candidate is \multiKKPiPi\%, \multiKPiPiPi\%, and
\multiPiPiPiPi\% for \DzToKKPiPi, \DzToKPiPiPi, and \DzToPiPiPiPi,
respectively.  For the signal mode with $\eta\to\pip\pim\piz$, 40\% of
events have multiple candidates. For $\eta\to\gamma\gamma$ and
$\omega$ decays, the number of events with multiple candidates is
$\sim10\%$, and for the remaining modes it is 1\% to 5\%.  If two or more
candidates are found in an event, the one with the highest
\Dstarp\ vertex \chisq\ probability is selected.  After applying the
best-candidate selection, the correct \Dstarp\ candidate in the
simulated samples is selected with a probability of 95\% or more for the
normalization modes. For the signal modes, 70\% of \Dstarp\ candidates
are correctly selected for $\eta\to\pip\pim\piz$, and between 86\% and
94\% for the remaining modes.  After the application of all selection
criteria and corrections for small differences between data and MC
simulation in tracking and PID performance, the reconstruction
efficiency $\epsilon_{\rm sig}$ for the simulated signal decays is
between 1.6\% and 3.6\%, depending on the mode. For the normalization
decays, the reconstruction efficiency $\epsilon_{\rm norm}$ is between
\effKKPiPiNoErr\% and \effPiPiPiPiNoErr\%. The difference between
$\epsilon_{\rm sig}$ and $\epsilon_{\rm norm}$ is mainly due to the
minimum momentum criterion on the leptons required by the PID
algorithms~\cite{TheBABAR:2013jta}.

\section{Signal Yield Extraction}
\label{fit}

The \DzToXzemu\ signal mode branching fraction $\BR_{\rm sig}$ is determined
relative to that of the normalization decay using

\begin{equation}
\label{eq:ratio}
  \BR_{\rm sig} = \frac{N_{\rm sig}}{N_{\rm norm}} 
  \frac{\epsilon_{\rm norm}}{\epsilon_{\rm sig}} \frac{\lum_{\rm
      norm}}{\lum_{\rm sig}} 
\frac{\BR_{\rm norm}}{\BR(X^0)},
\end{equation}

\noindent where \hbox{$\BR_{\rm norm}$} is the branching fraction of
the normalization mode~\cite{PDG2019}, and $N_{\rm sig}$ and $N_{\rm
  norm}$ are the fitted yields of the signal and normalization mode
decays, respectively. $\BR(X^0)$ is the branching
fraction of the intermediate meson decay channel. The symbols $\lum_{\rm
  sig}$ and $\lum_{\rm norm}$ represent the integrated luminosities of
the data samples used for the signal (\totallumi\invfb) and the
normalization decays (\usedlumi\invfb),
respectively~\cite{Lees:2013rw}. For the signal modes, we use both the
on-peak and off-peak data samples. For the normalization modes, a subset of the off-peak data is
sufficient for achieving statistical uncertainties that are much
smaller than the systematic uncertainties.

We perform an extended unbinned maximum likelihood fit to extract the
signal and background yields for both the normalization and signal
modes~\cite{Lees:2013gdj}. The likelihood function is 

\begin{equation}
\label{likelihood}
{\mathcal L} = \frac{1}{N!}\exp{\left(-\sum_{j=1}^{2}n_{j}\right)}
\prod_{i=1}^N\left[\sum_{j=1}^{2}n_{j}{\mathcal
    P}_{j}(\vec{x}_i;\vec{\alpha}_j)\right]\!.
\end{equation}

\noindent We define the likelihood for each event candidate $i$ to be the
sum of $n_j {\cal P}_j(\vec x_i; \vec \alpha_j)$ over two hypotheses
$j$ (signal or normalization and background). The symbol ${\cal
  P}_j(\vec x_i; \vec \alpha_j)$ is the product of the probability
density functions (PDFs) for hypothesis $j$ evaluated for the measured
variables $\vec x_i$ of the $i$-th event. The total number of events
in the sample is $N$, and $n_j$ is the yield for hypothesis $j$. The
quantities $\vec \alpha_j$ represent parameters of ${\cal P}_j$. The
distributions of each discriminating variable $x_i$ in the likelihood
function is modeled with one or more PDFs, where the parameters $\vec
\alpha_j$ are determined from fits to signal simulation or data samples.

Each normalization mode yield $N_{\rm norm}$ is extracted by
performing a two-dimensional unbinned maximum likelihood fit to the
\dm\ versus $m(\Dz)$ distributions in the range
$0.143<\dm<0.148\gevcc$ and $1.81<m(\Dz)<1.91\gevcc$. Considering
normalization and background events separately, the measured \dm\ and
$m(\Dz)$ values are essentially uncorrelated and are therefore treated
as independent observables in the fits. The PDFs in the fits depend on
the normalization mode and use sums of multiple
Cruijff~\cite{Lees:2018zz} and Crystal Ball~\cite{Skwarnicki:1986xj}
functions in both \dm\ and $m(\Dz)$. The functions for each observable
use a common mean. The background is modeled with an \argus\ threshold
function~\cite{Albrecht:1990am} for \dm\ and a Chebyshev polynomial
for $m(\Dz)$. The \argus\ end point parameter is fixed at
$0.1395\gevcc$, the \dm\ kinematic threshold for $\Dstarp\to\Dz\pip$
decays. All other PDF parameters, together with the normalization mode
and background yields, are allowed to vary in the fit. 

\begin{figure}[htbp!]
\begin{center}
  \begin{tabular}{c}
  \includegraphics[width=0.995\columnwidth]{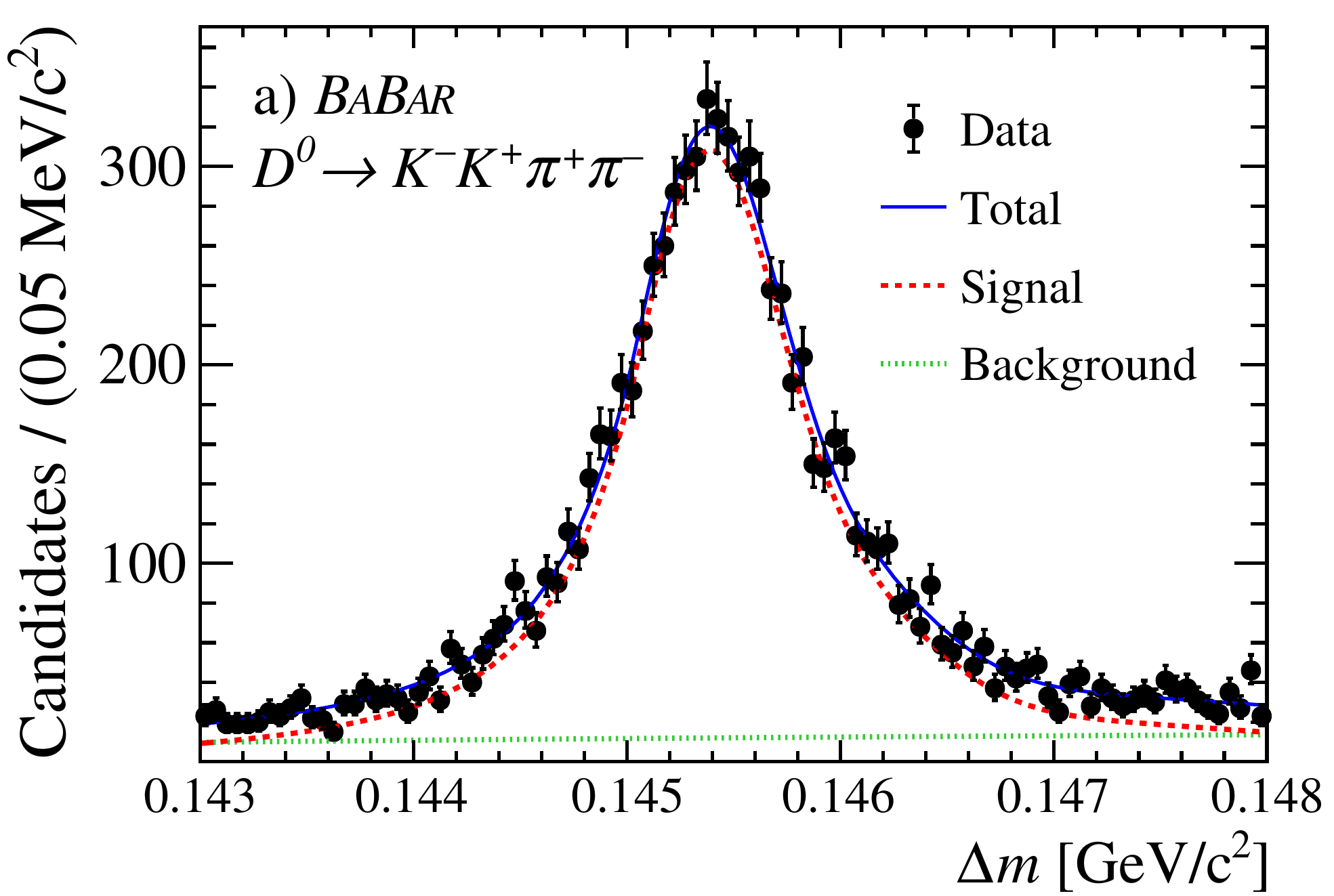} \\
  \includegraphics[width=0.995\columnwidth]{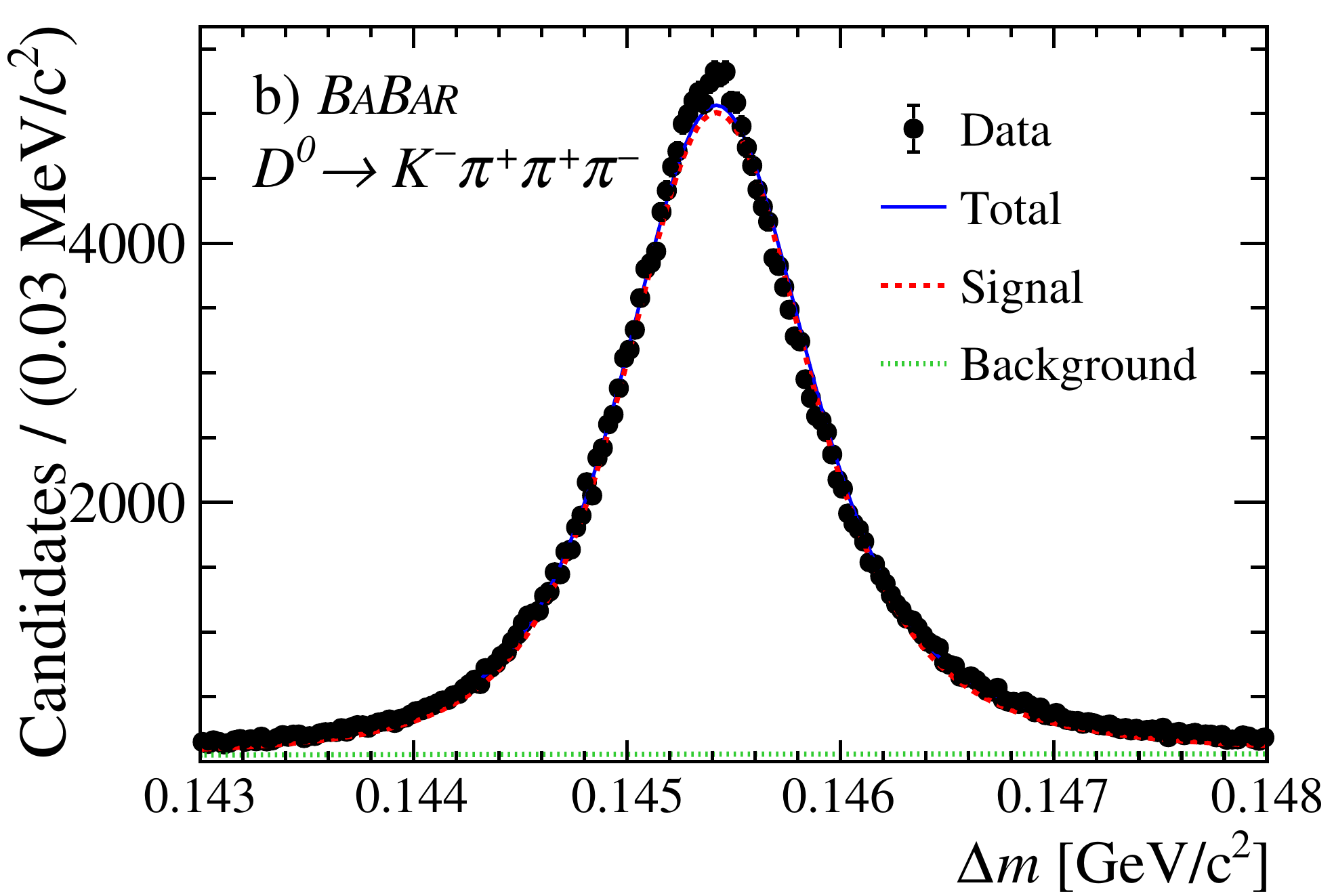} \\
  \includegraphics[width=0.995\columnwidth]{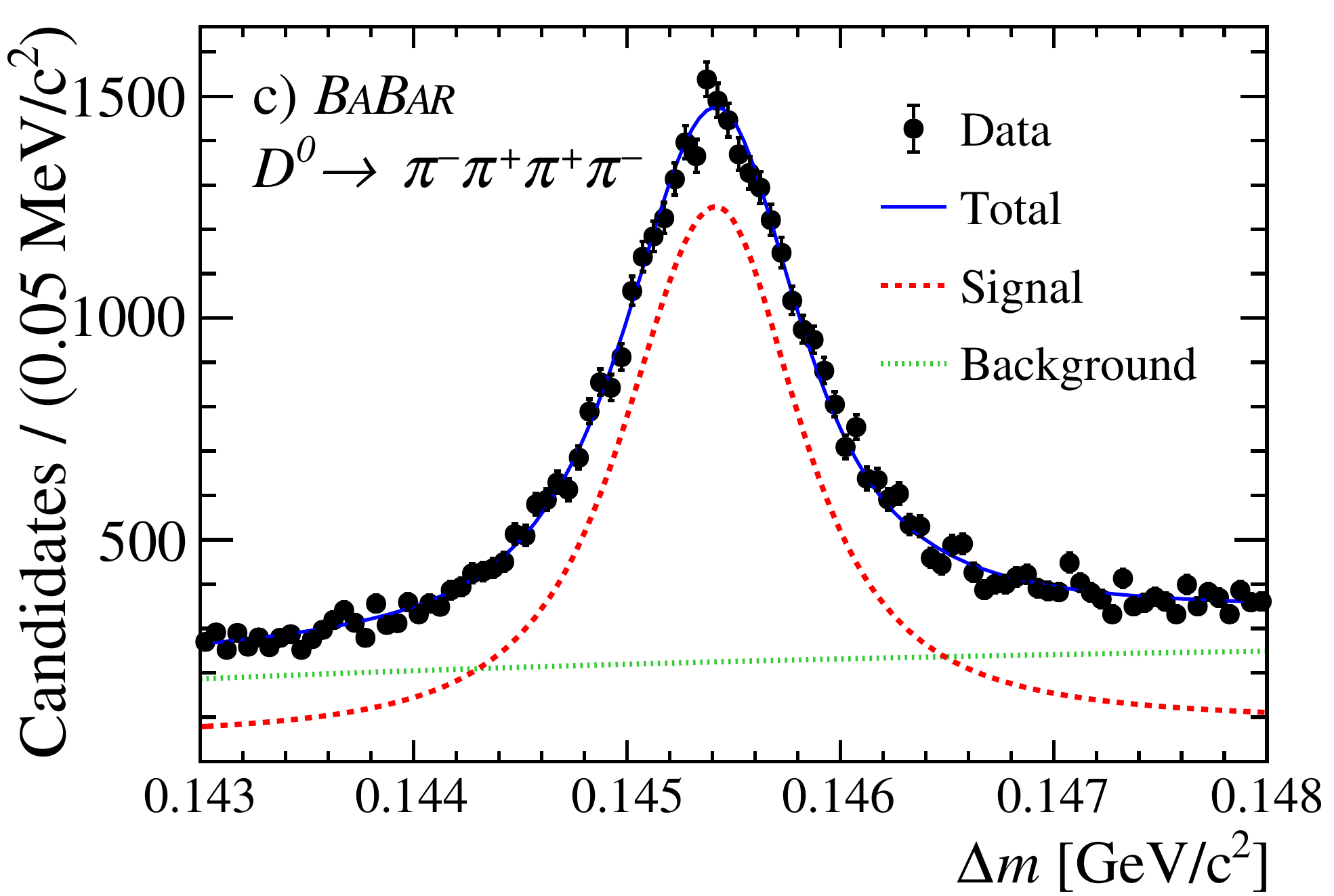}
\end{tabular}
\end{center}
\caption{Projections of the unbinned
  maximum-likelihood fits to the final candidate distributions as a function of \dm\ for the 
 normalization modes in the range $0.143<\dm<0.148\gevcc$. The solid
  blue line is the total fit, the dashed red line is the signal and
  the dotted green line is the background.}
\label{fig:fit_norm}
\end{figure}

The fitted yields and reconstruction efficiencies for the
normalization modes are given in
Table~\ref{tab:norm_fit}. Figure~\ref{fig:fit_norm} shows projections
of the unbinned maximum-likelihood fits onto the final candidate
distributions as a function of \dm\ for the normalization modes in the
range $0.143<\dm<0.148\gevcc$.

\begin{table}[htbp!]
\begin{center}
\caption[Normalization modes fitted yields]{Summary of fitted
  candidate yields, with statistical uncertainties, and reconstruction
  efficiencies for the three normalization modes.}
\begin{tabular}{lrc}
  \hline \hline \\ [-1em]
  Decay mode & \nnorm\ (candidates) & $\epsilon_{\rm norm}$ (\%)\\
   \hline
  \Dz\to\KPiPiPi  & \yieldKPiPiPi  &  \effKPiPiPi   \\
  \Dz\to\KKPiPi   & \yieldKKPiPi   &  \effKKPiPi       \\
  \Dz\to\PiPiPiPi & \yieldPiPiPiPi &  \effPiPiPiPi  \\
  \hline \hline
\end{tabular}
\label{tab:norm_fit}
\end{center}
\end{table}

\begin{figure}[htbp!]
\begin{center}
  \begin{tabular}{c}
  \includegraphics[width=0.995\columnwidth]{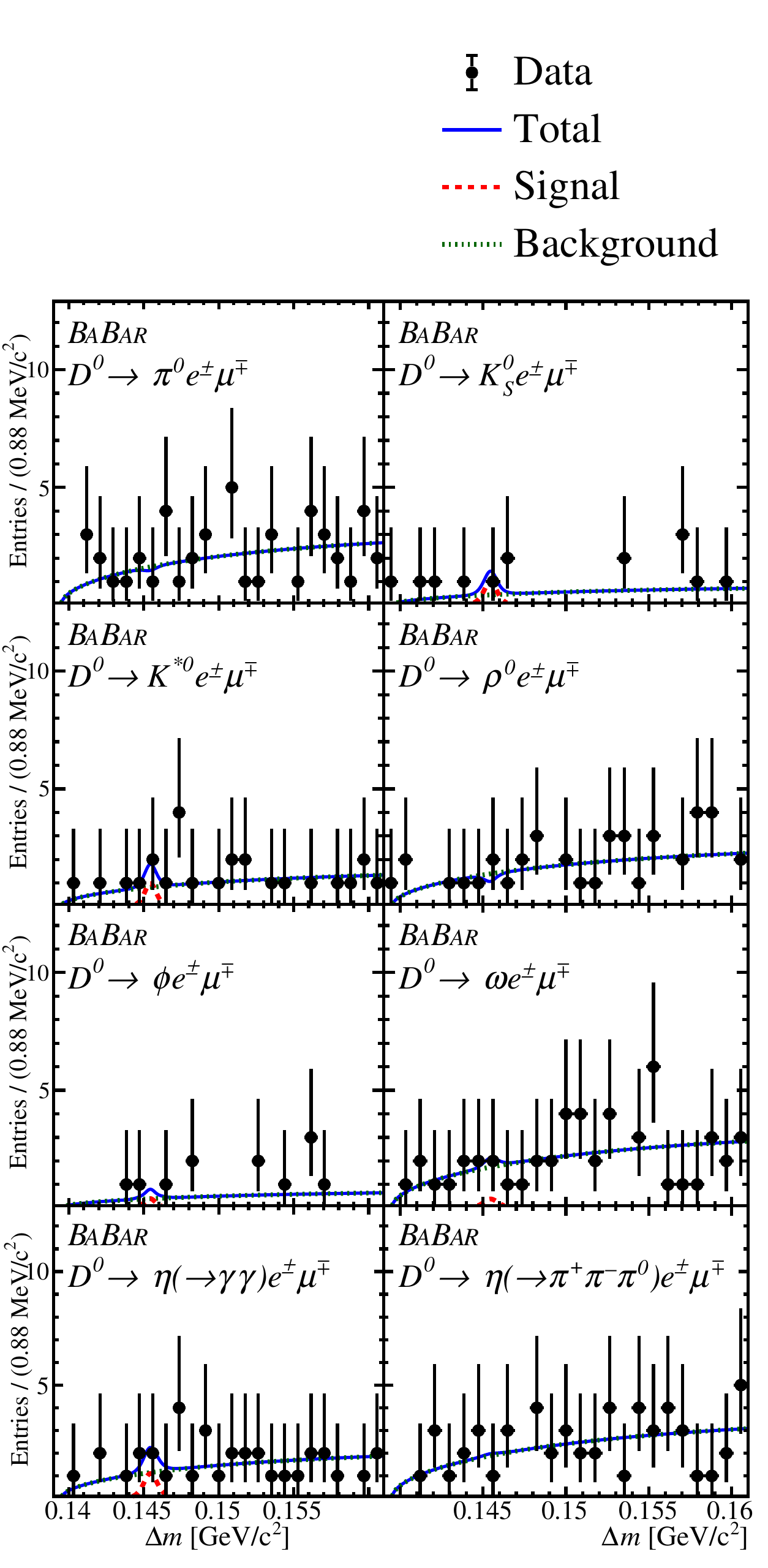}
\end{tabular}
\end{center}
\caption{Unbinned maximum-likelihood fits to the final candidate
  distributions as a function of \dm\ for the signal modes in the
  range $0.1395<\dm<0.1610\gevcc$. The solid blue line is the total
  fit, the dashed red line is the signal and the dotted green line is
  the background.}
\label{fig2}
\end{figure}

After the application of the selection criteria, there are on the
order of 100 events or fewer available for fitting in each signal
mode. Each signal mode yield \nsig\ is therefore extracted by
performing a one-dimensional unbinned maximum likelihood fit to
\dm\ in the range $0.1395<\dm<0.1610\gevcc$.  A Cruijff function is
implemented for the signal mode PDF, except for \DzToPhiemuOS, for which two
two-piece Gaussians functions are used, and \DzToRhoemuOS, for which two Cruijff
functions are used. The background is modeled with an \argus\ function with the
same end point used for the normalization modes. The signal PDF
parameters and the end point parameter are fixed in the fit. All other
background parameters and the signal and background yields are allowed
to vary. Figure~\ref{fig2} shows the results of the fits to the
\dm\ distributions for the signal modes.

We test the performance of the maximum likelihood fit for the
normalization modes by generating ensembles of MC samples from the
normalization and background PDF distributions. The mean numbers of
normalization and background candidates used in the ensembles are
taken from the fits to the data. The numbers of generated background
and normalization mode candidates are sampled from a Poisson
distribution. All background and normalization mode PDF parameters are
allowed to vary, except for the \argus\ function end point. No
significant biases are observed in the fitted yields of the normalization
modes. The same procedure is repeated for the maximum likelihood fits
to the signal modes, with ensembles of MC samples generated from the
background PDF distributions only, assuming a signal yield of zero.
The signal PDF parameters are fixed to the values used for the fits to
the data, and the signal yield is allowed to vary. The biases in the
fitted signal yields are less than $\pm0.3$ candidates for all modes,
and these are subtracted from the fitted yields before calculating the
signal branching fractions.

To confirm the normalization procedure, the signal modes in
Eq.~(\ref{eq:ratio}) are replaced with the decay \DzToKPi, which has a
well-measured branching fraction~\cite{PDG2019}. The \DzToKPi\ decays
are reconstructed using the on-peak data sample only
(\onreslumi\invfb). The \DzToKPi\ decay is selected using the same
criteria as used for the \DzToKPiPiPi\ mode, which is used as the
normalization mode for this test. The \mbox{\DzToKPi} signal yield is
\yieldKPi\ with $\epsilon_{\rm sig} = (\effKPi)\%$. Thus, we determine
$\BR(\DzToKPi) = (\BFDzToKPiPiPi)$\%, where the uncertainties are
statistical and systematic, respectively. This is consistent with the
current world average of $(\BFDzToKPi)$\%~\cite{PDG2019}. When the
test is repeated using either \DzToKKPiPi\ or \DzToPiPiPiPi\ as the
normalization mode, $\BR(\DzToKPi)$ is determined to be
$(\BFDzToKKPiPi)\%$ and $(\BFDzToPiPiPiPi)\%$, respectively.

\section{Systematic Uncertainties}
\label{syst}

The systematic uncertainties in the branching fraction determinations
of the signal modes arise from so-called additive systematic
uncertainties that affect the significance of the signal mode yields
in the fits to the data samples and from multiplicative
systematic uncertainties on the luminosity and signal reconstruction
efficiencies.

The main sources of the additive systematic uncertainties in the
signal yields are associated with the model parametrizations used in
the fits to the signal modes, the fit biases, the allowed
invariant-mass ranges for the \Dz\ and $X^{0}$ candidates, the amount
of cross feed, and the limited MC and data sample sizes available for
the optimization of the BDT discriminants.

The uncertainties associated with the fit model parametrizations of
the signal modes are estimated by repeating the fits with alternative
PDFs. This involves replacing the Cruijff functions with Crystal Ball
functions, using a two-piece Gaussian function, and changing the
number of functions used in the PDFs. For the background, the
\argus\ function is replaced by a first- or second-order
polynomial. The largest deviation occurs when using the Crystal Ball
functions for the signal and the first-order polynomial for the
background. The systematic uncertainty is taken as half this maximum
deviation. The largest contribution comes from the normalization mode
\DzToPiPiPiPi\ due to the presence of increased background and greater
uncertainty in the background shape. To account for potential
inaccuracies in the simulation of the \Dz\ and $X^0$ invariant mass
distributions, we change the mass selection ranges by $\pm 0.5\sigma$,
where $\sigma$ is the RMS width of the \Dz\ or $X^0$ meson.

The systematic uncertainties in the correction on the fit biases for
the signal yields are taken from the ensembles of fits to the MC
samples. Given the central value of the signal yield obtained from the fit in
each mode, the cross feed yields from all other modes are calculated and
are taken as a systematic uncertainty. To evaluate the systematic uncertainty
 in the application of the BDT discriminant, we vary the value of the selection
criterion for the BDT discriminant output, change the size of the
hidden region in data, and also retrain the BDT discriminant using a
training sample with a different ensemble of MC samples.  Summing the
uncertainties in quadrature, the total additive systematic
uncertainties in the signal yields are between 0.4 and 0.9 events.

Multiplicative systematic uncertainties are due to assumptions made
about the distributions of the final-state particles in the signal
simulation modeling, the model parametrizations used in the fits to
the normalization modes, the normalization mode branching fractions,
tracking and PID efficiencies, limited simulation sample sizes, and
luminosity.

Since the decay mechanism of the signal modes is unknown, we vary the
angular distributions of the simulated final-state particles from the
\Dz\ signal decay in three angular variables, defined following the
prescription of Ref.~\cite{Aubert:2004cp}. We weight the events, which
are simulated uniformly in phase space, using combinations of $\sin$,
$\cos$, $\sin^2$, and $\cos^2$ functions of the angular variables. The
reconstruction efficiencies calculated from simulation samples as
functions of the three angles are constant, within the statistics
available. The deviations of the reweighted efficiencies from the
default average reconstruction efficiencies are therefore small. Half
the maximum change in the average reconstruction efficiency is
assigned as a systematic uncertainty.

Uncertainties associated with the fit model parametrizations of the
normalization modes are estimated by repeating the fits with
alternative PDFs. This involves swapping the Cruijff and Crystal Ball
functions used in both \dm\ and $m(\Dz)$. For the background, the
order of the polynomials is changed and the \argus\ function is
replaced by a second-order polynomial. Half the maximum change in the
fitted yield is assigned as a systematic uncertainty.  The normalization modes branching fraction uncertainties are taken from
Ref.~\cite{PDG2019}.

For both signal and normalization modes, we include uncertainties to
account for discrepancies between reconstruction efficiencies
calculated from simulation and data samples of 1.0\% per \KS, 0.8\%
per lepton, and 0.7\% per hadron track~\cite{Allmendinger:2012ch}. We
include a momentum-dependent \piz\ reconstruction efficiency
uncertainty of 2.1\% for \DzToPizemuOS\ and 2.3\% for
\DzToOmegaemuOS\ and \DzToEtaThreepiemuOSsub. For the PID
efficiencies, we assign an uncertainty of 0.7\% per track for
electrons, 1.0\% for muons, 0.2\% for charged pions, and 1.1\% for
kaons~\cite{TheBABAR:2013jta}. A systematic uncertainty of 0.4\% is
associated with our knowledge of the luminosities $\lum_{\rm norm}$
and $\lum_{\rm sig}$~\cite{Lees:2013rw}. We assign systematic
uncertainties in the range 0.8\% to 1.8\% to account for the limited
size of the simulation samples available for calculating
reconstruction efficiencies for the signal and normalization modes.

The simulation samples for the normalization modes contain a resonant
structure of intermediate resonances that decay to two- or three-body
final states, as well as four-body nonresonant decays. To investigate
how changes in the resonant structure affect the reconstruction
efficiencies, the simulation samples were generated using a four-body
phase-space distribution only and the reconstruction efficiencies
recalculated. The resulting changes in reconstruction efficiencies are
less than the statistical uncertainties on $\epsilon_{\rm norm}$ due
to the limited size of the simulation samples, and no systematic
uncertainties are assigned. The total multiplicative systematic
uncertainties are between 4.7\% and 6.8\% for the normalization modes
and between 4.2\% and 7.8\% for the signal modes.

Table~\ref{tab:systematics_norm} summarizes the contributions of the
systematic uncertainties of the normalization modes to the systematic
uncertainties in the signal mode branching fractions, as defined in
Eq.~(\ref{eq:ratio}). Table~\ref{tab:systematics_signal} summarizes the
systematic uncertainties in the signal mode yields, excluding those
due to the normalization modes.

\begin{table}[htb!]
  \begin{center}
    \caption[Summary of normalization mode systematics]{Summary of the
      contributions to the systematic uncertainties on the signal mode
      branching fractions, as defined in Eq.~(\ref{eq:ratio}), that
      arise from uncertainties in the measurement of the normalization
      modes.}
    \begin{tabular}{rccc}
    \hline
    \hline
                     & \PiPiPiPi & \KPiPiPi & \KKPiPi\\
    \hline
    PDF variation          & 4.6\% & 1.0\% & 1.0\% \\
    \KS\ correction        & 1.0\% & 1.0\% & 1.0\% \\
    Tracking correction    & 3.5\% & 3.5\% & 3.5\% \\
    PID correction         & 0.8\% & 1.7\% & 2.6\% \\
    Luminosity             & 0.4\% & 0.4\% & 0.4\% \\
    Normalization \calB    & 3.0\% & 1.8\% & 4.5\% \\
    Simulation size        & 1.0\% & 1.0\% & 0.8\% \\
    \hline
   Total  & \systPiPiPiPi\% & \systKPiPiPi\%  & \systKKPiPi\% \\
\hline
    \hline
\end{tabular}
\label{tab:systematics_norm}
\end{center}
\end{table}

\begin{table*}[htb!]
  \begin{center}
    \caption[Summary of \DzToXzemu\ signal mode systematics]{Summary
      of \DzToXzemu\ additive and multiplicative systematic uncertainties, excluding
      those due to the normalization modes given in
      Table~\ref{tab:systematics_norm}.}
     \begin{tabular}{rcccccccc}
    \hline
    \hline \\ [-1em]
    $X^0 =$ & \piz & \KS & \Kstarzb & \rhoz & $\phi$ & $\omega$ & $\eta$ & $\eta$ \\
    $X^0\to$ & $\gamma\gamma$ & $\pip\pim$ & $\Km\pip$ &  $\pip\pim$ & $\Kp\Km$&  $\pip\pim\piz$ & $\gamma\gamma$ & $\pip\pim\piz$ \\
    
    \hline
    \multicolumn{2}{l}{Additive (events):} \\
    PDF variation        & 0.23 & 0.05 & 0.20 & 0.16 & 0.17 & 0.26 & 0.43 & 0.16 \\
    Fit bias             & 0.09 & 0.28 & 0.21 & 0.15 & 0.24 & 0.09 & 0.08 & 0.07 \\
    \Dz/$X^0$ mass       & 0.30 & 0.04 & 0.05 & 0.07 & 0.07 & 0.07 & 0.04 & 0.23\\
    BDT discriminant     & 0.83 & 0.68 & 0.71 & 0.30 & 0.06 & 0.35 & 0.27 & 0.58 \\
    Cross feed           &      &      &      & 0.01 &      &      & 0.06 & \\
    \hline
    Subtotal (candidates)  & 0.92 & 0.74 & 0.76 & 0.38 & 0.31 & 0.45 & 0.52 & 0.65 \\
    \hline
    \multicolumn{2}{l}{Multiplicative (\%):} \\
    Angular variation    & 1.4 & 2.8 & 2.0 & 3.4 & 5.3 & 1.9 & 1.6 & 1.6 \\
    $\calB(X^0)$ subdecay &     & 0.1 &     &     & 1.0 & 0.8 & 0.5 & 1.2 \\
    \KS\ correction       &     & 1.0  &  & & &  & &  \\
    Tracking correction   & 2.3 & 3.7 & 3.7 & 3.7 & 3.7 & 3.7 & 2.3 & 3.7 \\
    PID correction        & 2.7 & 2.1 & 3.0 & 2.1 & 3.9 & 3.1 & 2.7 & 3.1 \\
    \piz\ correction      & 2.1 &      &  & & & 2.3 & & 2.3 \\
    Luminosity            & 0.4 & 0.4 & 0.4 & 0.4 & 0.4 & 0.4 & 0.4 & 0.4 \\
    Simulation sample size        & 1.4 & 1.3 & 1.5 & 1.3 & 1.4 & 1.8 & 1.3 & 1.5 \\
   \hline
   Subtotal (\%)         & 4.2 & 5.4 & 5.4 & 5.6 & 7.8 & 5.7 & 4.2 & 5.6 \\
    \hline
    \hline
    \end{tabular}
   \label{tab:systematics_signal}
  \end{center}
\end{table*}

\section{Results}
\label{results}

Table~\ref{tab:hh_results} gives the fitted signal yields,
reconstruction efficiencies, branching fractions with statistical and
systematic uncertainties, 90\% C.L. upper limits on the branching
fractions, and previous upper
limits~\cite{Freyberger:1996it,Aitala:2000kk,PDG2019} for the signal
modes. The yields for all the signal modes are compatible with
zero. We assume that there are no cancellations due to
correlations in the systematic uncertainties in the numerator and
denominator of Eq.~(\ref{eq:ratio}). We use the frequentist approach
  of Feldman and Cousins~\cite{Feldman:1997qc} to determine 90\%
  C.L. bands.  When computing the limits, the systematic uncertainties
  are combined in quadrature with the statistical uncertainties in the fitted signal yields.

\begin{table*}[htb!]
\centering
\caption[Summary of results]{Summary of fitted signal yields \nsig\ with statistical and
  systematic uncertainties, reconstruction efficiencies $\epsilon_{\rm
  sig}$, branching fractions with statistical and systematic
  uncertainties, 90\% C.L. upper limits (U.L.) on the branching
  fractions, and previous
  limits~\cite{Freyberger:1996it,Aitala:2000kk,PDG2019}. The additive
  and multiplicative uncertainties are combined to obtain the overall
  systematic uncertainties. The branching fraction systematic
  uncertainties include the uncertainties in the normalization mode
  branching fractions.}
\begin{tabular}{lrccrrr}   
\hline \hline \\ [-1em]
& \multicolumn{1}{c}{\nsig} & \hspace{0.2cm}
&$\epsilon_{\rm sig}$ & \multicolumn{1}{c}{\BR $(\times 10^{-7})$}  &
\multicolumn{2}{c}{\BR\ 90\% U.L. $(\times 10^{-7})$}\\
 Decay mode & \multicolumn{1}{c}{(candidates)} & & (\%) &  & \babar\ & Previous\\
\hline \\ [-1em]
\DzToPizemuOS   &  \obsPizemuOS   &  & \effPizemuOS &
\bfactPizemuOS  & \ulactPizemuOS & 860\\
\DzToKSemuOS &  \obsKSemuOS   &  & \effKSemuOS & \bfactKSemuOS   & \ulactKSemuOS & 500\\
\DzToKstaremuOS &  \obsKstaremuOS   &  & \effKstaremuOS & \bfactKstaremuOS   & \ulactKstaremuOS & 830\\
\DzToRhoemuOS &  \obsRhoemuOS   &  & \effRhoemuOS & \bfactRhoemuOS   & \ulactRhoemuOS & 490\\
\DzToPhiemuOS &  \obsPhiemuOS &  & \effPhiemuOS & \bfactPhiemuOS   & \ulactPhiemuOS & 340\\
\DzToOmegaemuOS &  \obsOmegaemuOS &  & \effOmegaemuOS & \bfactOmegaemuOS & \ulactOmegaemuOS & 1200\\
\DzToEtaggemuOS &  &  &  & \bfactEtaemuOS & \ulactEtaemuOS & 1000\\
\hspace{0.5cm}with $\eta\to\gamma\gamma$  & \obsEtaggemuOS       &  & \effEtaggemuOS       & \bfactEtaggemuOS     & \ulactEtaggemuOS & \\
\hspace{0.5cm}with $\eta\to\pip\pim\piz$  & \obsEtaThreepiemuOS  &
& \effEtaThreepiemuOS  & \bfactEtaThreepiemuOS & \ulactEtaThreeemuOS & \\
\hline \hline
\end{tabular}
\label{tab:hh_results}
\end{table*}

In summary, we report 90\% C.L. upper limits on the branching
fractions for seven lepton-flavor-violating \mbox{\DzToXzemu}
decays. The analysis is based on a sample of $\epem$ annihilation data
collected with the \babar\ detector, corresponding to an integrated
luminosity of \totallumi\invfb. The limits are in the range
$(\ulactRhoemuOS - \ulactEtaemuOS)\times 10^{-7}$ and are between 1
and 2 orders of magnitude more stringent than previous
\DzToXzemu\ decay results. For the four \DzToXzemu\ decays with the
same final state as the \DzTohhemuOS\ decays reported in
Ref.~\cite{Lees:2019zz}, the limits are 1.5 to 3 times more stringent.

\section{Acknowledgments}
We are grateful for the 
extraordinary contributions of our \pep\ colleagues in
achieving the excellent luminosity and machine conditions
that have made this work possible.
The success of this project also relies critically on the 
expertise and dedication of the computing organizations that 
support \babar.
The collaborating institutions wish to thank 
SLAC for its support and the kind hospitality extended to them. 
This work is supported by the
US Department of Energy
and National Science Foundation, the
Natural Sciences and Engineering Research Council (Canada),
the Commissariat \`a l'Energie Atomique and
Institut National de Physique Nucl\'eaire et de Physique des Particules
(France), the
Bundesministerium f\"ur Bildung und Forschung and
Deutsche Forschungsgemeinschaft
(Germany), the
Istituto Nazionale di Fisica Nucleare (Italy),
the Foundation for Fundamental Research on Matter (The Netherlands),
the Research Council of Norway, the
Ministry of Education and Science of the Russian Federation, 
Ministerio de Econom\'{\i}a y Competitividad (Spain), the
Science and Technology Facilities Council (United Kingdom),
and the Binational Science Foundation (U.S.-Israel).
Individuals have received support from 
the Marie-Curie IEF program (European Union) and the A. P. Sloan Foundation (USA).

\end{document}